\documentclass[11pt]{article}

\usepackage{graphicx}
\usepackage{epstopdf}
\usepackage{amsmath}
\usepackage{amssymb}
\usepackage{amsfonts}
\usepackage{amsthm}
\usepackage[usenames]{color}
\usepackage{array}

\usepackage[
colorlinks=true,
linkcolor=blue,
urlcolor=blue,
filecolor=blue,
citecolor=red,
pdfstartview=FitV,
pdftitle={},
pdfauthor={},
pdfsubject={},
pdfkeywords={},
pdfpagemode=None,
bookmarksopen=true
]{hyperref}

\usepackage{epsfig}

\usepackage{hyperref}

\newcommand{\bea}{\begin{eqnarray}}
\newcommand{\eea}{\end{eqnarray}}
\newcommand{\ba}{\begin{array}}
	\newcommand{\ea}{\end{array}}
\newcommand{\ee}{\end{equation}}

\numberwithin{equation}{section}

\begin{document}

\begin{flushright}
	\texttt{\today}
\end{flushright}

\begin{centering}
	
	\vspace{2cm}
	
	\textbf{\Large{
			  First Law of Entanglement Entropy in   Flat-Space Holography }}
	
	\vspace{0.8cm}
	
	{\large   Reza Fareghbal, Mehdi Hakami Shalamzari }
	
	\vspace{0.5cm}
	
	\begin{minipage}{.9\textwidth}\small
		\begin{center}
			
			{\it  Department of Physics, 
				Shahid Beheshti University, 
				G.C., Evin, Tehran 19839, Iran.  }\\

			\vspace{0.5cm}
			{\tt  r$\_$fareghbal@sbu.ac.ir, reza.fareghbal@gmail.com }
			\\ \tt $ \, $ma$\_$hakami@sbu.ac.ir \\
			
		\end{center}
	\end{minipage}


	\begin{abstract}
	According to flat/Bondi-Metzner-Sachs invariant field theories (BMSFT)	correspondence, asymptotically flat spacetimes in $(d+1)$-dimensions are  dual to $d$-dimensional BMSFTs. In this duality, similar to the Ryu-Takayanagi proposal in the AdS/CFT correspondence, the entanglement entropy of subsystems in the field theory side is given by the area of some particular surfaces in the gravity side. In this paper we find the holographic counterpart of the first law of entanglement entropy (FLEE)  in a two-dimensional BMSFT. We show that FLEE for the BMSFT perturbed states which are descried by three-dimensional flat-space cosmology,  corresponds to the integral of a particular one-form on a closed curve. This curve consists of BMSFT interval and also  null and spacelike geodesics in the bulk gravitational theory. Exterior derivative of this form is zero when it is calculated for the flat-space cosmology. However, for a generic perturbation of three-dimensional global Minkowski spacetime, the exterior derivative of one-form yields Einstein equation.   This is the first step for constructing bulk geometry by using FLEE in the flat/BMSFT correspondence.  
	\end{abstract}

\end{centering}

\newpage

\tableofcontents

\section{Introduction}
Flat/BMSFT is an extension of AdS/CFT correspondence to  non-AdS geometries. According to this duality quantum gravity in the asymptotically flat spacetimes in $(d+1)$-dimensions can be described by a $d$-dimensional field theory which is BMS-invariant \cite{Bagchi:2010zz,Bagchi:2012cy}. In the gravity side,  BMS symmetry is the asymptotic symmetry of asymptotically flat spacetimes at null infinity \cite{BMS,Ashtekar:1996cd}. In  the field theory side , the global part of  BMS algebra is given by ultra-relativistic contraction of conformal algebra . Thus one can interpret the flat-space limit (zero cosmological constant limit) in the gravity side as the  ultrarelativistic limit of CFT in the boundary theory \cite{Bagchi:2012cy}. In this view, one can study flat/BMSFT  by starting from AdS/CFT and taking a limit, the flat-space limit in the bulk and the ultrarelativistic limit in the boundary.   

BMS symmetry as the asymptotic symmetry, is infinite-dimensional in three and four dimensions \cite{Barnich:2006av}-\cite{aspects}. Hence one may expect to find some universal aspects for two- and three- dimensional BMSFTs. This situation is very similar to the two-dimensional conformal field theories (CFTs) which their infinite-dimensional symmetry is used to predict the structure of correlation  functions  as well as entanglement entropy of subsystems. Similarly, the entanglement entropy formula for some particular intervals in BMSFT$_2$ has been introduced in \cite{Bagchi:2014iea} by just using the infinite symmetry of two-dimensional BMSFTs and then studied more carefully in \cite{Hosseini:2015uba}-\cite{Asadi:2018lzr}. 

In the context of AdS/CFT correspondence, the entanglement entropy of CFT subsystems has a holographic description. According to Ryu-Takayanagi proposal, this entropy is proportional to the area of a bulk surface which has the minimum area among the surfaces  connected to the  boundary subsystem \cite{Ryu:2006bv,Hubeny:2007xt}. A similar proposal for the BMSFT entanglement entropy has been introduced in \cite{Jiang:2017ecm}. Accordingly, the BMSFT entanglement entropy can be given by the area of particular surfaces. These surfaces are not connected directly  to the boundary of subsystem but there are  null rays  which connect them to null infinity where the subsystem is supposed to live. The corresponding surface, null rays and the subsystem together construct a closed surface . 

Another interesting problem which was studied in the context of AdS/CFT is the holographic description of the first law of the entanglement entropy (FLEE). It was shown in \cite{Lashkari:2013koa,Faulkner:2013ica} that writing both sides of  FLEE in terms of corresponding bulk parameters finally yields linearized Einstein equations. In other words, FLEE as a constraint in the boundary theory reduces to a constraint on the bulk geometry  which is exactly Einstein equation. If this connection is an intrinsic  property of gauge/gravity dualities, one can use entanglement entropy and its first law in an arbitrary field theory to find a dual gravitational geometry. 

In this paper we study the proposal of \cite{Lashkari:2013koa,Faulkner:2013ica}  in the context of flat$_3$/BMSFT$_2$ correspondence. We start from FLEE and use flat/BMSFT correspondence  to write it in terms of  components of the  asymptotically flat  bulk metric. We focus on the BMSFT states which their gravitational  dual are flat-space cosmology \cite{Cornalba:2002fi}-\cite{Barnich:2012xq}. It is shown that both sides of the FLEE formula can be written in terms of the integral of an one-form over curves consist of BMSFT interval and the null and the spacelike geodesics introduced in \cite{Jiang:2017ecm}. These curves construct a closed curve, thus  one can use  Stokes's theorem to write integrals as the integral of the external derivative of the one-form over the surface bounded by the curves. For the metric of the flat-space cosmology, the exterior derivative of this form is zero. For a generic metric which satisfy BMS boundary condition (see for example \cite{Lambert:2014poa}), the exterior derivative of one-form results in Einstein equation.    Our work is not only the first step  generalization of the proposal of \cite{Lashkari:2013koa,Faulkner:2013ica} for the flat-space holography but also shows that the flat/BMSFT correspondence studied in several previous works (see references in \cite{Grumiller:2019xna}) is a worthwhile  duality.

In section two we  review the proposal of \cite{Faulkner:2013ica} in the context of AdS/CFT. In section three after briefly reviewing the flat/BMSFT correspondence and holographic description of BMSFT entanglement entropy, we write  FLEE in terms of bulk metric  and deduce the Einstein equation.   

\section{Linear bulk equation from FLEE in AdS/CFT}
\subsection{Entanglement entropy and its first law}\label{sec2}
For a quantum field theory   state $|\psi \rangle$, the density matrix is 
\begin{equation}
\rho = |\psi\rangle \langle \psi |.
\end{equation}
If we decompose a spatial (time constant) slice $\Sigma$ to two  subsystems $B$ and $\bar{B}$ $(\Sigma=B\cup \bar{B})$, then 
the density matrix associated to $B$ can be obtained from $\rho$ by tracing out the degrees of freedom of the complement subsystem $\bar{B}$ as
\begin{equation}
\rho_B = \text{tr}_{\bar{B}} \rho.
\end{equation}
The Entanglement entropy of subsystems $B$ is  the von Neumann entropy associated to the density matrix $\rho_B$,
\begin{equation}\label{ENTanglement entropy}
S_B = -\text{tr}(\rho_B \ln \rho_B).
\end{equation}

For a small perturbation $|\psi(\varepsilon) \rangle$ to the initial state  $|\psi(0) \rangle$ of the whole system, the first law of entanglement entropy (FLEE) is 
\begin{equation}\label{FLEE1}
\delta S_B =\frac{d}{d \varepsilon} S_B =\frac{d}{d \varepsilon} \langle H_B \rangle = \frac{d}{d \varepsilon} \text{tr}\left(H_B\rho_B\right) \equiv \delta E_B,
\end{equation}
where $H_B$ is  modular Hamiltonian which is independent of perturbation and  defined through
\begin{equation}
H_B =- \ln {\rho_B}\left(\epsilon=0\right).
\end{equation}
 Formula \eqref{FLEE1} is a quantum generalization of the first law of thermodynamics. This formula holds for any arbitrary small perturbation of quantum state and for any subsystem $B$.
 
 Mostly, it is difficult  to compute the modular Hamiltonian $H_B$ and  its associated density matrix $\rho_B$. However, for the cases that $H_B$ is a local operator, one  may  find a unitary transformation (and hence reversible which acts also on the corrdinates)   which maps $\rho_B$ to a thermal density matrix. Hence the resulatant entropy   is a thermal one (see \cite{Casini:2011kv}). If we denote the unitary transformation by $U$ and the final thermal  density matrix by $\rho_{\mathcal{H}}$, then
\begin{equation}\label{rhobh}
\rho_B = U \rho_{\mathcal{H}} U^{-1}.
\end{equation}
It is not difficult to check that the thermal entropy given by 
\begin{equation}
S_{TH}=-\text{tr}( \rho_{\mathcal{H}}\ln \rho_{\mathcal{H}}),
\end{equation}
is the same as the entanglement entropy \eqref{ENTanglement entropy}.
Since $\rho_{\mathcal{H}}$ is thermal, it can be written as\footnote{We have absorbed a factor of $2\pi$ into the definition of $H_{\mathcal{H}}$ }. 
\begin{equation}
\rho_{\mathcal{H}}=\frac{e^{- H_{\mathcal{H}}}}{\text{tr}\left(e^{- H_{\mathcal{H}}}\right)},
\end{equation}
where $H_{\mathcal{H}}$ is the associated charge of the  symmetry generator $\xi$. $\xi$ is called modular flow and generates translation along the thermal circle of the transformed coordinates. Thus firstly one  can  apply this unitary transformation and calculate the  thermal entropy with the help of $H_{\mathcal{H}}$ and then through the inverse unitary transformation \eqref{rhobh} calculate the density matrix $\rho_B$ (or equivalently modular Hamiltonian $H_B$). Moreover, it is clear that $H_B$ is the conserved charge of $\xi$ upto an additive constant. This constant can be ignored when the variation of the modular Hamiltonian in  FLEE is considered. In the rest of this paper we mostly use  modular flow instead of modular Hamiltonian.

\subsection{Holographic FLEE in AdS/CFT}

Formula \eqref{FLEE1} holds for small perturbations in any quantum field theory. One may ask about the holographic counterpart of this formula for the field theories which have holographic duals. The first step is applying FLEE for the CFTs and wondering about the holographic formula in the dual AdS geometry in the context of AdS/CFT. It was shown in \cite{Lashkari:2013koa, Faulkner:2013ica} that FLEE for a CFT yields the linearized equations of motion in the AdS gravity side.
 In this subsection  we review the derivation. 

Suppose a $d-$dimensional CFT on Minkowski spacetime $\mathbb{R} ^{1, d-1}$. The dual  $(d+1)-$dimensional holographic  dual consists the   asymptotically AdS spacetimes. For the vacuum state the dual spacetime  is pure AdS whose metric $g_{ab}^0$ in the Feffermann-Graham coordinates reads
\begin{equation}
ds^2= \frac{\ell^2}{z^2}(\eta_{\mu \nu}dx^{\mu}dx^{\nu}+dz^2),
\end{equation}

 We consider  a spacial time slice $\Sigma$ of $d-$dimensional Minkowski space  and divide it to two regions $B$ and $\bar{B}$ ($\Sigma=B \cup \bar{B}$). Let B be a $(d-1)-$ dimensional ball with radius $R$. 
                   
In order to find $\delta E_B$ in \eqref{FLEE1},  we need to calculate the  vacuum expectation value of the modular Hamiltonian.
The modular Hamiltonian for this ball shaped region is calculated in \cite{Casini:2011kv} as follows
\begin{equation}
H_B =2 \pi \int_B \mathrm{d}^{d-1}\,x \, \frac{R^2-\delta_{ij} (x^i -x_0 ^i)(x^j -x_0 ^j)}{2R} \ T^{tt}(x),
\end{equation}
 where $x_0^i$ are the coordinates of the center of ball $B$ and  $T^{\mu\nu}$ is the stress tensor of CFT.  We use the convention $ x^\mu=(t, x^i)$.
Hence FLEE \eqref{FLEE1} can be written as
\begin{equation}
\delta S_B  =2 \pi \int_B \mathrm{d}^{d-1}\,x \ \frac{R^2-\delta_{ij} (x^i -x_0 ^i)(x^j -x_0 ^j)}{2R} \  \delta \langle T^{tt}(x) \rangle.
\end{equation}
 Now  we use holography to calculate $\delta S_B$. When the CFT vacuum state $|\Psi(0) \rangle$ is perturbed to the state $|\Psi(\varepsilon) \rangle$, in the dual gravitational theory,   the metric of the dual AdS spacetime  will be perturbed as
\begin{equation}
ds^2= \frac{\ell^2}{z^2} ((\eta_{\mu \nu}+h_{\mu \nu})dx^{\mu}dx^{\nu}+dz^2),
\end{equation}
where $h_{\mu \nu}$ are infinitesimal. 
By means of the Ryu-Takayanagi formula \cite{Ryu:2006bv,Hubeny:2007xt}  we can write
\begin{equation}
S_B = S_{HEE} = \frac{A_{\tilde{B}}}{4G},
\end{equation}
where 
 $A_{\tilde{B}}$ is  the minimal  area of the co-dimension two surface $\tilde{B}$ in the bulk AdS space which  is homologous to $B$ and given by
\begin{equation}
A_{\tilde{B}} = \int_{\tilde{B}} \mathrm{d}^{d-1}\sigma \sqrt{det(\gamma_{AB})}.
\end{equation}
 Here $\gamma_{AB}$ is the induced metric on  $\tilde{B}$. 

Let us illustrate  the  holographic counterpart  of $\delta S_B$ and $\delta E_B $ ,respectively, as $\delta S_B^{grav.}$ and $\delta E_B^{grav.}$. It was shown in \cite{Lashkari:2013koa,Faulkner:2013ica} that they are given  as follows in terms of bulk perturbed metric $h_{ij}$:
\begin{equation}
\delta S_B^{grav.} = \frac{\ell^{d-3}}{8 G R} \int _{\tilde{B}} \mathrm{d} ^{d-1}\,x \,\big(R^2 \delta^{ij}-(x^i x^j)\big)\, h_{ij}(x,z),
\end{equation}
\begin{equation}
\delta E_B^{grav.} = \frac{\ell^{d-3} d}{16 G R} \int _{{B}} \mathrm{d} ^{d-1}\,x \,\big(R^2 -(\vec{x}- \vec{x}_0)^2\big)\, \delta^{ij}\, h_{ij}(x,z=0).
\end{equation}
Thus the FLEE formula \eqref{FLEE1} is written as
\begin{equation}\label{FLEE integral}
 \int _{\tilde{B}} \mathrm{d} ^{d-1}\,x \,\big(R^2 \delta^{ij}-(x^i x^j)\big)\, h_{ij}=\frac{ d}{2} \int _{{B}} \mathrm{d} ^{d-1}\,x \,\big(R^2 -(\vec{x}- \vec{x}_0)^2\big)\, \delta^{ij}\, h_{ij}.
\end{equation}
This is a non-local equation which is correct for any ball shaped region with arbitrary  radius $R$ and  center coordinate $\{x_0^i\}$. Thus one may think about a local equation which is equivalent to \eqref{FLEE integral}. In order to find this local constraint, we look for a form $\chi$ such that 
\begin{equation}
\int_B \chi = \delta E_{B}^{grav.}, \qquad \int_{\tilde{B}} \chi =\delta S_{{B}}^{grav.}.
\end{equation}
If such a form $\chi$ exists, using  \eqref{FLEE1} we can write
\begin{equation}\label{chitodchi}
\delta S_B^{grav.} - \delta E_B^{grav.}=0=\int_{\tilde{B}} \chi -\int_B \chi =\int_{B \cup \tilde{B}} \chi =\int_{\Pi} \mathrm{d}\chi,
\end{equation}
where $\Pi$ is the hypersurface  bounded by $B$ and $\tilde{B}$ $(B \cup \tilde{B} = \partial \Pi)$ and located at $t=t_0$.
For the asymptotically AdS spacetimes, $\chi$ is given by \cite{Faulkner:2013ica}
\begin{equation}\label{chi1}
\chi = -\, \frac{1}{16 \pi G} \left[ \delta \left(\nabla^{a}\xi^{b} \epsilon_{ab}  \right)+\xi^b \epsilon_{ab}\left(\nabla _c h^{ac}\ - \nabla ^a h^c _c \right)\right],
\end{equation}
where $\xi^a$ is the bulk modular flow
\begin{equation}
\xi = - \frac{2 \pi}{R} (t-t_0)\left[z \partial_z +(x^i-x^i_0)\partial_i \right] + \frac{\pi}{R} \left[ R^2 -z^2 -(x^i-x_0^i)^2-(t-t_0)^2\right]\partial_t.
\end{equation}
For this form, the exterior derivative is given by
\begin{equation}\label{dxidg0}
\mathrm{d} \chi = -\frac{1}{8 \pi G}\xi^a \delta G_{ab}\epsilon^{b},
\end{equation}
where $\delta G_{ab}$ are linearized Einstein equations around  AdS spacetimes,
\begin{align}
\delta G_{ab}=&-\tfrac12 \nabla_b \nabla_a h^c{}_c+\tfrac12 \nabla_c \nabla_a h_b{}^c+\tfrac12 \nabla_c \nabla_b h_a{}^c-\tfrac12 \nabla_c \nabla^c h_{ab}-\tfrac12 g_{ab} \nabla_d \nabla_c h^{cd}+\tfrac12 g_{ab} \nabla_d \nabla^d h^c{}_c\notag \\ &-\frac{2\Lambda}{d-1}\left(h_{ab}-\tfrac12 g_{ab} h^c{}_c\right)
\end{align}
  and $\epsilon ^b $ is related to volume form as follows
\begin{equation}\label{epsilon}
\epsilon ^a = g^{ab} \frac{1}{d\,!}\, \bar{\epsilon} _{bi_2...i_{d+1}} \sqrt{-g}\  \mathrm{d}x^{i_2} \wedge \dots \wedge \, \mathrm{d}x^{i_{d+1}}.
\end{equation}
Moreover, the exterior derivative is zero on the boundary. 

From \eqref{chitodchi} and \eqref{dxidg0} it is obvious that the holographic interpretation of the first law of entanglement entropy leads to
\begin{equation}\label{FLEE int local}
\int_{\Pi} \xi^a \delta G_{ab}\epsilon^{b} = 0.
\end{equation}
Using the fact that only the $t$ component of $\xi^a$ is non-vanishing on $\Pi$ and also FLEE is valid for all of the ball shaped regions with arbitrary $R$,  from \eqref{FLEE int local} one can deduce that \cite{Jaksland:2017nqx} 
\begin{equation}
 \delta G_{tt} = 0.
 \end{equation}
In the above derivation, $B$ was a constant time slice in the boundary. Thus for a constant time slices or rest frame of references, we can deduce the $tt$ component of the linearized Einstein equation. Repeating  the same argument for the ball shaped regions in the arbitrary frame of references we can find $\delta G_{\mu\nu} = 0$ where $\mu$ and $\nu$ are directions of the field theory. Moreover, from the fact that exterior derivative of $\chi$ is zero on the boundary we can deduce that $\delta G_{z\mu} = 0$ and $\delta G_{zz} = 0$ on the boundary or $z=0$. Thus  all component of the linearized Einstein equation are zero at $z=0$. One can use this result as the initial condition and using the Bianchi identity prove that $\delta G_{z\mu}$ and $\delta G_{zz}$ are zero everywhere \cite{Wald84}. 
 
We see that the  gravitational interpretation of  FLEE  in  CFTs leads to the linearized equations of motion of the dual AdS gravity. In the next section  we will apply the above procedure for  asymptotically flat spacetimes in the context of  flat/BMSFT correspondence.
\section{Holographic FLEE in Flat/BMSFT correspondence}
\subsection{Flat/BMSFT correspondence}
 Asymptotic symmetries of the asymptotically AdS spacetimes in $(d+1)$ dimensions  are the same as local symmetries of the $d-$dimensional CFTs. One may expect such an equivalence between  the gravity solutions and their dual field theory    for the non-AdS spacetimes. Asymptotically AdS spacetimes are solutions of Einstein gravity with negative cosmological constant. Taking the flat space limit which is equivalent to the zero cosmological constant limit results in asymptotically flat spacetimes. Although this limit is not well-defined for the asymptotically AdS spacetimes written in the Fefferman-Graham coordinate but it is possible to find appropriate coordinates with well-defined flat space limit \cite{Barnich:2012aw,Fareghbal:2013ifa}. A relevant question is finding a counterpart for the flat space limit of the gravity theory in the field theory side. To answer this question one needs to study the asymptotic symmetry of the asymptotically flat spacetimes. This study has been done in \cite{BMS} for the four dimensional  and in \cite{Ashtekar:1996cd} for the three dimensional spacetimes. More recent studies show that for the four dimensional cases the asymptotic symmetry algebra at null infinity is the semi-direct sum of infinite dimensional local conformal symmetry algebra on a two-sphere and the abelian ideal algebra of supertranslations \cite{Barnich:4dBMS}. This algebra is known as $bms_4$. Such an infinite dimensional locally well-defined symmetry algebra also exists at null infinity of  three dimensional asymptotically flat spacetimes \cite{Barnich:2006av} . This algebra is called $bms_3$. 
  
  The observation of \cite{Bagchi:2012cy} is that the $bms_3$ is isomorphic to an infinite-dimensional algebra in two dimensions which is given by ultra-relativistic contraction of conformal algebra. Thus it was proposed in \cite{Bagchi:2012cy} that the holographic dual of asymptotically flat spacetimes in $(d+1)$ dimensions are field theories in $d$ dimensions which have BMS symmetry. We call these BMS invariant field theories BMSFT and the correspondence between them and asymptotically flat spacetimes flat/BMSFT. 
  
  To be more precise, let us consider  Einstein-Hilbert action with negative cosmological constant in three dimensions
\begin{equation}
S = \frac{1}{16 \pi G} \int \mathrm{d}^3 x\, \sqrt{-g}\,(R+\frac{4}{\ell^2}).
\end{equation}
An appropriate coordinate with well-defined flat space limit is BMS gauge \cite{Barnich:2012aw}
\begin{equation}\label{BMS gauge}
ds^2 = \left( - \frac{r^2}{\ell^2} + \mathcal{M}\right)du^2 -2du dr +2 \mathcal{N}du d\phi +r^2 d\phi^2,
\end{equation} 
where  $\mathcal{M}$ and $\mathcal{N}$ are functions of $u$ and $\phi$ and are constrained  by using the equations of motion as 
\begin{equation}
\partial_u \mathcal{M} = \frac{2}{\ell^2} \partial_{\phi}\mathcal{N} \ , \qquad 2 \partial_u \mathcal{N} = \partial_{\phi} \mathcal{M}. 
\end{equation}
The asymptotic symmetry algebra is exactly the conformal algebra in two dimensions,
\begin{equation}
\begin{split}
&[\mathcal{L}_m , \mathcal{L}_n] =(m-n) \mathcal{L}_{m+n}, \\
&[\bar{\mathcal{L}}_m , \bar{\mathcal{L}}_n] =(m-n) \bar{\mathcal{L}}_{m+n},
\\
&[\mathcal{L}_m , \bar{\mathcal{L}}_n] = 0, \qquad\qquad\qquad\qquad m,n \in \mathbb{Z}.
\end{split}
\end{equation}
 The algebra of conserved charges is centrally extended with central charges $c=\bar c=3\ell/2G$.

Taking the flat space limit from metric \eqref{BMS gauge} yields asymptotically flat spacetimes with metric 
\begin{equation}\label{afm}
ds^2 = {M}du^2 -2du dr +2 {N}du d\phi +r^2 d\phi^2,
\end{equation}
where ${M}$ and ${N}$ are functions of $u$ and $\phi$ and they satisfy 
\begin{equation}
\partial_u {M} = 0 \ , \qquad 2 \partial_u {N} = \partial_{\phi} {M}, 
\end{equation}
The asymptotic symmetry algebra at null infinity is infinite dimensional $bms_3$ algebra \cite{Barnich:2006av},
\begin{equation}\label{bmsa}
\begin{split}
&[L_m , L_n] =(m-n) L_{m+n}, \\
&[L_m , M_n] =(m-n) M_{m+n},
\\
&[M_m , M_n] = 0, \qquad\qquad\qquad m,n \in \mathbb{Z}.
\end{split}
\end{equation}
 The algebra of conserved charges is also centrally extended.

The generators of $bms_3$ can be obtained by taking flat space limit from the generators of conformal algebra \cite{Barnich:2012aw},
\begin{equation}\label{flat limit of generators}
L_m = \lim\limits_{\frac{G}{\ell}\rightarrow 0} \, (\mathcal{L}_m-\bar{\mathcal{L}}_{-m}), \qquad M_m = \frac{G}{\ell} \lim\limits_{\frac{G}{\ell}\rightarrow 0} \, (\mathcal{L}_m+\bar{\mathcal{L}}_{-m}).
\end{equation}
It was argued in \cite{Bagchi:2012cy} that the limit \eqref{flat limit of generators} which is taken in the gravity side corresponds to the ultra relativistic limit in the field theory side. In the rest of this paper by  BMSFT$_2$ we mean a field theory which has the symmetry algebra \eqref{bmsa}.

From BMSFT$_3$ we mean a field theory with the following symmetry algebra
\begin{equation}
\begin{split}
&\left[ L_m, L_n\right] =(m-n)L_{m+n}, \\
&\left[\bar{L}_m, \bar{L}_n\right] =(m-n) \bar{L}_{m+n}, \\
&\left[ L_m, \bar{L}_n \right] =0, \\
&\left[L_l, M_{m,n} \right] =\left(\frac{l+1}{2}- m\right) M_{m+l,n},\\
&\left[\bar{L}_l, M_{m,n} \right] =\left(\frac{l+1}{2}- n\right) M_{m,n+l}, \qquad m,n,l \in \mathbb{Z}.
\end{split}
\end{equation}
This algebra is called $bms_4$  and is the asymptotic symmetry of the four dimensional asymptotically flat spacetimes  at null infinity \cite{Barnich:4dBMS}. $L_m$ and $\bar{L}_m$ are generators of super-rotations and $M_{m,n}$ are generators of  super-translations. The Poincare subalgebra is generated by
\begin{equation}
\left\lbrace L_{-1} , L_0 , L_{+1} , \bar{L}_{-1} , \bar{L}_0 , \bar{L}_{+1}, M_{0,0}, M_{0,+1}, M_{+1,0}, M_{+1,+1} \right\rbrace.
\end{equation}

\subsection{Holographic entanglement entropy in flat/BMSFT}
Similar to other field theories, it is possible to define entanglement entropy for the subsystems of BMSFT. The infinite dimensional symmetry of BMSFTs admits to find universal formulas for the entanglement entropy of sub-regions \cite{Bagchi:2014iea}. Moreover, using the flat/BMSFT correspondence one can find a holographic description for the BMSFT entanglement entropy. Recently, a prescription (similar to the   Ryu-Takayanagi's proposal for the CFT entanglement entropy \cite{Ryu:2006bv,Hubeny:2007xt})  has been proposed for the BMSFT entanglememnt entropy  \cite{Jiang:2017ecm} that relates  it to the area of some particular curves into the bulk flat spacetimes.
 According to \cite{Jiang:2017ecm}, the  entanglement entropy of sub-region $B$ of BMSFT$_2$ is given by
\begin{equation}\label{song formula}
S_{HEE} = \frac{\text{Length}(\gamma)}{4G} = \frac{\text{Length}(\gamma \cup \gamma_+ \cup \gamma_-)}{4G}
\end{equation}
where $\gamma$ is a spacelike geodesic and $\gamma_+$ and $\gamma_-$ are null rays from $\partial\gamma$ to $\partial B$.  

The most generic solution of Einstein gravity with zero cosmological constant in three dimensions is given by \eqref{afm}.  In the rest of this paper we will consider an interval $B$ in the BMSFT which is determined by $-\frac{l_u}{2}<u<\frac{l_u}{2}$ and $-\frac{l_{\phi}}{2}<\phi<\frac{l_{\phi}}{2}$ where $l_u$ and $l_{\phi}$ are constants. Among the various values of functions $M$ and $N$ in \eqref{afm}, the following three metrics are of more interest:
\begin{enumerate}
\item Null-orbifold or Poincar\'e patch  with metric (${M}= {N}=0$ in \eqref{afm})
\begin{equation}
ds^2=-2dudr+r^2 d\phi^2.  
\end{equation}
In this case the bulk modular flow is 
\begin{eqnarray}
\nonumber\xi^{bulk}= -\frac{\pi}{2 l_{\phi}} \Bigg[ \left(\l_{\phi}^2 - 4 \phi ^2 + \frac{8(u l_{\phi}-l_u \phi)}{r l_{\phi}}\right)\partial_{\phi} &+\left(l_u l_{\phi}+\frac{4l_u}{l_{\phi}}\phi^2 -8 u \phi\right)\partial_u\\&+\left(\frac{8l_u}{l_{\phi}}+8r\phi \right)\partial_r\Bigg].
\end{eqnarray}
 Here $\gamma$ is given by
\begin{equation}
r=-\frac{l_u}{l_{\phi}\phi}, \qquad u = \frac{l_u l_{\phi}}{8 \phi}+\frac{l_u \phi}{2 l_{\phi}}.
\end{equation}

By using the coordinate transformations 
\begin{equation}
\begin{split}
&t=\frac{l_{\phi}}{4} r + \frac{2}{l_{\phi}} u +\frac{1}{l_\phi} r \phi^2,  \\
&x=\frac{l_u}{l_\phi} + r \phi, \\
&y=\frac{l_{\phi}}{4} r -\frac{2}{l_{\phi}} u -\frac{1}{l_\phi} r \phi^2, 
\end{split}
\end{equation}
we can change the metric of null-orbifold to the Cartesian coordinate
\begin{equation}
ds^2 = -dt^2+dx^2+dy^2.
\end{equation}
In this  coordinates the bulk modular flow is given by
\begin{equation}\label{xicart1}
\xi^{bulk} = -2 \pi (x \partial_t + t \partial_x),
\end{equation}
and geodesics are 
\begin{equation}
\gamma : x=t=0, \qquad -\frac{l_u}{l_{\phi}} \le y \le + \frac{l_u}{l_{\phi}},
\end{equation}
\begin{equation}
\gamma_+ : x=t, \qquad y=-\frac{l_u}{l_{\phi}}, 
\end{equation}
\begin{equation}
\gamma_- : x=-t, \qquad y=+\frac{l_u}{l_{\phi}}.
\end{equation}

\item Global Minkowski with metric (${M}=-1$ and  ${N}=0 $ in \eqref{afm})
\begin{equation}\label{global Minkowski}
ds^2=-du^2-2dudr+r^2 d\phi^2.
\end{equation}

 The bulk modular flow is 
\begin{align}\label{bulk modular flow global}
\xi^{bulk} = &\pi \csc{\frac{l_{\phi}}{2}}\left(2(\cos\frac{l_\phi}{2}-\cos\phi)+\frac{1}{r}(l_u\sin\phi\cot\frac{l_\phi}{2}-2u\cos\phi)\right) \partial_{\phi}
\notag, \\
&+\pi \csc{\frac{l_{\phi}}{2}} \left(-l_u \csc{\frac{l_{\phi}}{2}}+l_u\cos{\phi}\cot{\frac{l_{\phi}}{2}}+2u \sin{\phi}\right)\partial_u \notag, \\
&-\pi \csc{\frac{l_{\phi}}{2}}\left(l_u \cos{\phi}\cot{\frac{l_{\phi}}{2}}+2(r+u)\sin{\phi}\right)\partial_r,
\end{align}
  where $\gamma$ is given by\footnote{We assume that $l_\phi<\pi$.}
\begin{equation}\label{gammaforglobal}
\qquad r=-\frac{l_u \csc{\frac{l_{\phi}}{2}}}{2 \sin{\phi}},\qquad u = -\frac{l_u}{2} \cot{\frac{l_{\phi}}{2}}\cot{\phi}-r.
\end{equation}

Using coordinate transformation \cite{Wen:2018mev}
\begin{equation}\label{globaltocartes}
\begin{split}
&t=(r+u)\csc{\frac{l_{\phi}}{2}}-r\cos{\phi} \cot{\frac{l_{\phi}}{2}}, \\
&x=r \sin{\phi} +\frac{l_u}{2} \csc{\frac{l_{\phi}}{2}}, \\
&y=r\cos{\phi} \csc{\frac{l_{\phi}}{2}} - (r+u)\cot{\frac{l_{\phi}}{2}}
\end{split}
\end{equation}
we have
\begin{equation}\label{cartesi}
ds^2= -dt^2+dx^2+dy^2.
\end{equation}
 In this Cartesian coordinates the bulk modular flow is the same as \eqref{xicart1}
and geodesics  are 
\begin{equation}\label{gamma1}
\gamma : x=0=t, \qquad -\frac{l_u}{2}\cot{\frac{l_{\phi}}{2}}\le y \le +\frac{l_u}{2}\cot{\frac{l_{\phi}}{2}},
\end{equation}
\begin{equation}\label{gammam}
\gamma_+ : x=t, \qquad y=-\frac{l_u}{2}\cot{\frac{l_{\phi}}{2}},
\end{equation}
\begin{equation}\label{gammam1}
\gamma_- : x=-t, \qquad y=+\frac{l_u}{2}\cot{\frac{l_{\phi}}{2}}.
\end{equation}
 
\item Flat-space cosmology (FSC) with metric ($M=m$ and $N=j$ )
\begin{equation}\label{FSC}
ds^2=mdu^2-2dudr+2jdud\phi+r^2d\phi^2,
\end{equation}
where $m$ and $j$ are constants. It has a cosmological horizon at radius $r_c=\frac{j}{\sqrt{m}}$. FSC is a shift-boost orbifold of Minkowski spacetime \cite{Cornalba:2003kd} and can be brought into the Cartesian coordinate locally by using the following transformation:
\begin{equation}\label{FSC to Cart}
\begin{split}
&r=\sqrt{m(t^2-x^2)+r_c^2},\\
&\phi=-\dfrac{1}{\sqrt{m}}\log\dfrac{\sqrt{m}(t-x)}{r+r_c},\\
&u=\dfrac{1}{m}\left(r-\sqrt{m}y-\sqrt{m}r_c\phi\right).
\end{split}
\end{equation}
\end{enumerate}

Both of the null-orbifold and global Minkowski  correspond to the BMSFT states which are non-thermal but for the null-orbifold, BMSFT is on a plane and for the global Minkowski the corresponding BMSFT is on the cylinder. FSC \eqref{FSC} corresponds to the BMSFT thermal states. The holographic  entanglement entropy of interval $B$  is given by
\begin{equation}\label{HEE FSC}
S=\dfrac{1}{2G}\left[\dfrac{\pi}{\beta_\phi}\left(l_u+\dfrac{\beta_u}{\beta_\phi}\ell_\phi\right)\coth\left(\dfrac{\pi\ell_\phi}{\beta_\phi}\right)-\dfrac{\beta_u}{\beta_\phi}\right],
\end{equation}
where 
\begin{equation}\label{def beta}
\beta_\phi=\dfrac{2\pi}{\sqrt{m}},\qquad \dfrac{\beta_u}{\beta_\phi}=\dfrac{j}{m}.
\end{equation}
\subsection{Holographic FLEE}\label{sec5}
In this section we will consider the BMSFT  dual to the global Minkowski.
The starting point is FLEE formula \eqref{FLEE1} which is written in the field theory side. We want to use Flat$_3$/BMSFT$_2$ to write both sides of this formula in the gravity side.  BMSFT lives on a cylinder with coordinates $(u,\phi)$ and  interval $B$  is given by $-\frac{l_u}{2}<u-u_0<\frac{l_u}{2}$ and $-\frac{l_\phi}{2}<\phi-\phi_0<\frac{l_\phi}{2}$ where $l_u$, $l_\phi$, $u_0$ and $\phi_0$ are constants.

Let us start from the right hand side of \eqref{FLEE1}. In order to calculate the expectation value of modular Hamiltonian, we use the fact that up to an additive constant, the modular Hamiltonian $H_B$ is the same as conserved charge of the modular flow $\xi$. If we show the stress tensor of BMSFT by $T_{ab}$, the corresponding charge of $\xi$ can be calculated on a spacelike surface  $\Sigma$ with  metric  $\sigma_{ab}$ as \cite{Brown:1992br}
\begin{equation}\label{def of charge bmsft}
  Q_{\xi}=\int_\Sigma d\sigma \sqrt{\text{det}\left(\sigma_{ab}\right)}n_a\xi^b T^a_b,
  \end{equation}  
where $\sigma$ is the coordinate on the surface $\Sigma$ and $n^a$ is the unit timelike vector normal to $\Sigma$. The most challenging problem in the flat-space holography is definition of $\Sigma$. In the AdS/CFT correspondence, $\Sigma$ is a spacelike ( surface on  the conformal boundary of the asymptotically AdS spacetimes. However, such a definition for   conformal infinity of asymptotically flat spacetimes is not appropriate in the flat-space holography . In the previous works \cite{Fareghbal:2013ifa},\cite{Fareghbal:2014oba}-\cite{Fareghbal:2018xii}, in the flat-space holography,   $\Sigma$ has been defined by using the corresponding surface of asymptotically AdS spacetimes which their flat-space limit yields the asymptotically flat metric.  To be precise, let us consider AdS$_3$ metric written in the BMS coordinate,
\begin{equation}\label{ads bms}
ds^2=-\left(1+\dfrac{r^2}{\ell^2}\right)du^2-2dud\phi+r^2
d\phi^2.
\end{equation}
where $\ell$ is the radius of AdS space. At fixed but large $r$ we can write,
\begin{equation}
ds_B^2=\dfrac{r^2}{\ell^2}\left(- du^2+ \ell^2d\phi^2\right)+\mathcal{O}(r^0),
\end{equation}
Thus we can write the metric of conformal boundary as
\begin{equation}\label{CB2}
ds_{CB}^2=- du^2+ \ell^2d\phi^2.
\end{equation}
In the AdS/CFT correspondence, the metric of $\Sigma$ in \eqref{def of charge bmsft} is given by using \eqref{CB2}.  The new point in all of papers \cite{Fareghbal:2013ifa},\cite{Fareghbal:2014oba}-\cite{Fareghbal:2018xii} is that \eqref{CB2} is also appropriate for writing   metric of $\Sigma$ in the $\ell\to\infty$  limit. The proposal of \cite{Fareghbal:2013ifa}  for the definition of $\Sigma$ is that we use a metric similar to \eqref{CB2} but replace $\ell$ with three dimensional Newton constant $G$. In this paper we employ this definition  of  $\Sigma$. Since we want to study  FLEE in a BMSFT which is holographic dual of global Minkowski, the metric of bulk spacetime is given by 
\eqref{global Minkowski} which is the $\ell\to\infty$ limit of \eqref{ads bms}. Thus we choose $\Sigma$ as a spacelike subspace of a space which is determined by metric
\begin{equation}\label{CB flat}
ds_{CB}^2=- du^2+ G^2d\phi^2.
\end{equation}

 It will prove convinient to first make a coordinate transformation as
\begin{equation}
w=u-u_0-\dfrac{l_u}{2}\dfrac{\sin(\phi-\phi_0)}{\sin\frac{l_\phi}{2}}.
\end{equation}
In this coordinate, our interval will be on the $\phi$ axe between $-\frac{l_\phi}{2}<\phi-\phi_0<\frac{l_\phi}{2}$. Moreover, by taking $r\to\infty$ limit from \eqref{bulk modular flow global}, we can find the BMSFT modular flow on the interwal $(w=0)$ as
\begin{equation}\label{vector in w}
 \xi^w=0,\qquad \xi^\phi=\dfrac{2\pi}{\sin\frac{l_\phi}{2}}\left(\cos\frac{l_\phi}{2}-\cos(\phi-\phi_0)\right).
 \end{equation} 
If we determine $\Sigma$ as $w=0, -\frac{l_\phi}{2}<\phi-\phi_0<\frac{l_\phi}{2}$ then using \eqref{CB flat} and \eqref{vector in w} we find
\begin{equation}\label{RHS1}
\delta E_B=\delta \langle H_B \rangle=\dfrac{2\pi G}{\sin\frac{l_\phi}{2}}\int_{\phi_0-\frac{l_\phi}{2}}^{\phi_0+\frac{\l_\phi}{ 2}}d\phi\,\left(\cos\frac{l_\phi}{2}-\cos(\phi-\phi_0)\right)\delta\langle T_\phi^w\rangle
\end{equation}
Hence we can write the write hand side of \eqref{FLEE1} in terms of BMSFT stress tensor by using flat-space holography.  

In order to calculate the left hand side of \eqref{FLEE1} holographically, we perturb the metric of global coordinate \eqref{global Minkowski} as 
\begin{equation}\label{perturbed metric}
ds^2=(-1+h_{uu})du^2-2 du dr+2h_{u\phi}du d\phi+r^2d\phi^2.
\end{equation}
We consider the case which  $h_{uu}$ and $h_{u\phi}$ are constants. With this choice \eqref{perturbed metric} is similar to flat-space cosmology \eqref{FSC}.
  For writing \eqref{perturbed metric} we do not use equations of motion. The fixed  components of metric have been determined by using boundary conditions which are necessary to have BMS symmetry at  null infinity (see for example \cite{Lambert:2014poa}). In other words, the fact that the dual theory is BMSFT imposes \eqref{perturbed metric} for the form of metric. This is similar to choosing Fefferman-Graham coordinate in the context of AdS/CFT correspondence. Line element \eqref{perturbed metric} is not the generic one which fulfils the BMS boundary conditions. In order to simplify equations we have fixed some components. However, our argument in the rest of paper can be  generalized to  more generic cases.

Since $h_{uu}$ and $h_{u\phi}$ are infinitesimal constants, we can use \eqref{HEE FSC} to calculate $\delta S$. We find
\begin{equation}\label{LHS1fixed h}
\delta S={1\over 4G}\left[2\left(-1+{l_\phi\over2}\cot{{l_\phi\over2}}\right)h_{u\phi}+{l_u\over2}\left(\cot{l_\phi\over2}-{l_\phi\over2\sin^2{l_\phi\over2}}\right)h_{uu}\right].
\end{equation}
Using \eqref{RHS1} and \eqref{LHS1fixed h}, we can write the FLEE as 
\begin{equation}\label{EFLEE for fixed h}
\begin{split}
 \int_{\phi_0-\frac{l_\phi}{2}}^{\phi_0+\frac{\l_\phi}{ 2}}d\phi\,\left(\cos\frac{l_\phi}{2}-\cos(\phi-\phi_0)\right)&\delta\langle T_\phi^w\rangle=\cr&\!\!\!\!\!\!\!\!\!\!{\sin{l_\phi\over2}\over 8\pi G^2}\left[2\left(-1+{l_\phi\over2}\cot{{l_\phi\over2}}\right)h_{u\phi}+{l_u\over2}\left(\cot{l_\phi\over2}-{l_\phi\over2\sin^2{l_\phi\over2}}\right)h_{uu}\right].
\end{split}
 \end{equation} 
This formula is valid for all of intervals determined by $l_\phi$, $l_u$ and $(u_0,\phi_0)$. For a very small interval which is given by $l_\phi\to 0$, $l_u\to 0$ but ${l_u\over l_\phi}=$fixed, the expectation value of stress tensor can be considered as a function of center of the interval. Since center of interval is an arbitrary point, using \eqref{EFLEE for fixed h} we find,
\begin{equation}\label{stress for fixed h}
 \delta\langle T_\phi^w\rangle={1\over 8\pi G^2}\left(h_{u\phi}+{l_u\over 2}{\cos\phi\over\sin{l_\phi\over 2}}h_{uu}\right).
 \end{equation} 
Putting \eqref{stress for fixed h} into \eqref{RHS1}, we find $\delta E_B$ as
\begin{equation}\label{final dE for fixed h}
\delta E_B=\dfrac{1}{4G\sin\frac{l_\phi}{2}}\int_{\phi_0-\frac{l_\phi}{2}}^{\phi_0+\frac{\l_\phi}{ 2}}d\phi\,\left(\cos\frac{l_\phi}{2}-\cos(\phi-\phi_0)\right)\left(h_{u\phi}+{l_u\over 2}{\cos\phi\over\sin{l_\phi\over 2}}h_{uu}\right).
\end{equation}

The interesting point is that both of $\delta S_B$ and $\delta E_B$ given   by \eqref{LHS1fixed h} and \eqref{final dE for fixed h} are written as the integral of a specific one-form $\chi$. Precisely, we can write\footnote{ In the global Minkowski coordinate, $\gamma_+$ consists of two null curves connected at $r=0$ \cite{Jiang:2017ecm}. Since $\chi$ is singular at $r=0$, we use  contour  $r=\epsilon$ in the calculation of $\int_{\gamma_+}\!\!\!\! \chi$ and after integration take $\epsilon\to 0$. }
\begin{equation}
\delta E=\int_B \chi,\qquad \delta S=\int_{\gamma_-\cup\gamma\cup\gamma_+}\!\!\!\!\!\!\!\!\!\!\!\!\!\!\!\!\!\!\!\!\chi,
\end{equation}
where
\begin{equation}\label{form 2}
\chi={1\over 16\pi G}\epsilon_{\mu\nu\alpha}\left[\xi^\nu\nabla^\mu h-\xi^\nu\nabla_\sigma h^{\mu\sigma}+\xi_\sigma\nabla^\nu h^{\mu\sigma}+{1\over2} h\nabla ^\nu\xi^\mu+{1\over2}h^{\nu\sigma}\left(\nabla^\mu\xi_\sigma-\nabla_\sigma\xi^\mu\right)\right]dx^\alpha.
\end{equation}
$\xi$ is the bulk modular flow \eqref{bulk modular flow global}, $h=h_\mu^\nu$ and $\epsilon_{\mu\nu\alpha}$ is the completely antisymmetric tensor with component $\epsilon_{012}=\sqrt{|g_0|}$ where $g_0$ is the determinant of global Minkowski \eqref{global Minkowski}. Thus the FLEE formula \eqref{FLEE1} for BMSFT can be written as
\begin{equation}\label{FLEE form}
\int_B \chi-\int_{\gamma_-\cup\gamma\cup\gamma_+}\!\!\!\!\!\!\!\!\!\!\!\!\!\!\!\!\!\!\!\!\chi\quad=0.
\end{equation}
  Curves $B$ and $\gamma_-\cup\gamma\cup\gamma_+$ construct a closed  path. Hence, we can write \eqref{FLEE form} as 
\begin{equation}\label{FLEE dform integral}
 \int_\Pi d\chi=0,
 \end{equation} 
where $d\chi$ is the exterior derivative of $\chi$ and $\Pi$ is any surface bounded by $B\cup\gamma_-\cup\gamma\cup\gamma_+$. Since $\Pi$ is any bounded surface, from \eqref{FLEE dform integral} one may expect  that 
\begin{equation}\label{FLEE final} 
d\chi=0.
\end{equation}
It is not difficult to check  that \eqref{FLEE final} is satisfied for the perturbed metric given by \eqref{perturbed metric}. In fact, the metric \eqref{perturbed metric} with constant $h_{uu}$ and $h_{u\phi}$ is a solution of Einstein equation. 

Let us consider a case which $h_{uu}$ and $h_{u\phi}$ are arbitrary functions of $u$ and $\phi$. Now we have
\begin{equation}\label{dchi}
d\chi={1\over 16  G r^2} \left(d\chi_{ru }dr \wedge du+d\chi_{u \phi } du\wedge d \phi \right),
\end{equation}
where
\begin{equation}\label{dchiru}
d\chi_{ru}=  \left(\partial_\phi h_{uu}-2\partial_u h_{u\phi}\right)\left(l_u\cot{l_\phi\over2}\cos{\theta}-l_u \csc{l_\phi \over2}+2u \sin{\theta}\right),
\end{equation}
and
\begin{align}\label{dchiuphi}
d\chi_{u \phi}=\nonumber & r\left\{\left(\partial_\phi h_{uu}-2\partial_u h_{u\phi}\right)\left[-\cot{l_\phi \over2}\left(2r+l_u \csc{l_\phi \over2}\sin{\theta}\right)+2\cos{\theta}\csc{l_\phi \over2}\left(r+u\right)\right]\right\}
\\ \nonumber
& +r \csc{l_\phi \over2}\partial_{\phi}\left(\partial_\phi h_{uu}-2\partial_u h_{u\phi}\right)\left(l_u\cot{l_\phi\over2}\cos{\theta}-l_u \csc{l_\phi \over2}+2u \sin{\theta}\right)
\\
&+ r^2 \csc{l_\phi \over2} \partial_u h_{uu} \left(l_u\cot{l_\phi\over2}\cos{\theta}-l_u \csc{l_\phi \over2}+2u \sin{\theta}\right),
\end{align}
Thus using  \eqref{FLEE final}, \eqref{dchiru} and \eqref{dchiuphi} we find that
\begin{equation}\label{final equations}
\partial_\phi h_{uu}=2\partial_u h_{u\phi},\qquad \partial_u h_{uu}=0.
\end{equation}  
These are the relation which one can conclude from the Einstein equation for the metric \eqref{perturbed metric}.

\section{Summary and Conclusion} 
In this paper we studied another aspect of flat/BMSFT which was previously introduced in the context of AdS/CFT. We wrote  FLEE of  BMSFT$_2$  in terms of  three-dimensional asymptotically flat metrics. The steps are analogue to those that are used in the context  of  AdS/CFT correspondence. We rewrite both sides of FLEE \eqref{FLEE1} by using corresponding bulk parameters. $\delta S_B$ in \eqref{FLEE1}  is the  variation  of entanglement entropy with respect to the state by which the system is described. Using the proposal of \cite{Jiang:2017ecm} one can write this variation  as the   variation of length  of  some spatial curves in the bulk geometry. $\delta E_B$ in  the right hand side of FLEE \eqref{FLEE1} is variation of the expectation value of the modular Hamiltonian. For calculating  this quantity, we used  the fact that the modular Hamiltonian is the conserved charge of modular flow upto an additive constant which can be ignored in the variation. BMSFT  conserved charges are given by using stress tensor. Using flat/BMSFT dictionary  we relate the calculation of the conserved charges  to a bulk calculation similar to the Brown-York proposal \cite{Brown:1992br}. The keypoint in this calculation is the definition of the spatial surface over  which the integration is performed. In the AdS/CFT correspondence this surface is given by using the conformal boundary of asymptotically AdS spacetimes. In this case we do not use the standard definition of conformal boundary. Our proposal is that this surface for the flat spacetimes is the same as that one for the asymptotically AdS case whose  flat-space limit yields the asymptotically flat spacetimes \cite{Fareghbal:2013ifa}. This proposal works again in this problem similar to all previous works \cite{Fareghbal:2014oba}-\cite{Fareghbal:2018xii}, however, a thorough investigation is necessary that  we hope to do in our future studies.

In this paper we assumed that the perturbed state in the field theory side corresponds to  a metric similar to the flat-space cosmology \cite{Cornalba:2002fi}-\cite{Barnich:2012xq} in the bulk theory. Hence, the gravitational counterpart of FLEE was the exterior derivative   of a one-form which is zero for the flat-space cosmology. The exterior derivative of this form for a generic metric which satisfy BMS boundary condition results in Einstein equations for undermined components of the metric. This is a good hint that holographic FLEE  is Einstein equation in the flat/BMSFT correspondence. \\

\textbf{Note added:} While we were ready to submit this work, ref. \cite{Godet:2019wje} was posted on the arXiv whose results overlap with ours.

\subsubsection*{Acknowledgements}
The authors would like to thank Seyed Morteza Hosseini and Pedram Karimi for useful comments and discussions. R.F. is grateful for the hospitality of CERN theory department where some part of current paper was done.   This work is supported by  Iran National Science Foundation (INSF), project No. 97017212.

\end{document}